\documentclass[pdflatex,sn-mathphys-num]{sn-jnl}

\usepackage{graphicx}%
\usepackage{multirow}%
\usepackage{amsmath,amssymb,amsfonts}%
\usepackage{amsthm}%
\usepackage{mathrsfs}%
\usepackage[title]{appendix}%
\usepackage{xcolor}%
\usepackage{textcomp}%
\usepackage{manyfoot}%
\usepackage{booktabs}%
\usepackage{algorithm}%
\usepackage{algorithmicx}%
\usepackage{algpseudocode}%
\usepackage{listings}%
\usepackage{physics}

\newcommand{\review}[1]{\textcolor{black}{#1}}

\theoremstyle{thmstyleone}%
\theoremstyle{thmstyletwo}%

\theoremstyle{thmstylethree}%

\begin{document}

\title[Article Title]{(1+1)-Dimensional Schr\"odinger-Poisson equation with contact interaction}


\author{\fnm{\'Oscar} \sur{Rodr\'iguez-Villalba}}\email{oscareduardo.rodriguezvillalba@unipr.it}

\author{\fnm{Ivan} \sur{Saychenko}}\email{ivan.saychenko@unipr.it}

\author{\fnm{Massimo} \sur{Pietroni}}\email{massimo.pietroni@unipr.it}

\author*{\fnm{Sandro} \sur{Wimberger}}\email{sandromarcel.wimberger@unipr.it}

\affil{\orgdiv{Dipartamento di Scienze Matematiche, Fisiche e Informatiche}, \orgname{Universit\`a di Parma}, \orgaddress{\street{Campus Universitario, Parco Area delle Scienze n. 7/a}, \city{Parma}, \postcode{43124}, \country{Italy}}}

\affil{\orgdiv{INFN}, \orgname{Sezione di Milano Bicocca, Gruppo Collegato di Parma}, \orgaddress{\street{Campus Universitario, Parco Area delle Scienze n. 7/a}, \city{Parma}, \postcode{43124}, \country{Italy}}}



\abstract{We investigate the role of contact interactions in the dynamics of fuzzy dark matter (FDM) modeled through the Schr\"odinger-Poisson equation in one spatial dimension. While the $\Lambda$CDM paradigm successfully explains structure formation on large scales, its small-scale predictions remain in tension with observations. FDM offers an alternative framework, where local self-interactions can further influence the formation and evolution of structures. We explore both attractive and repulsive contact interactions in static and expanding backgrounds. Using numerical simulations, we examine their impact on three key scenarios: the properties of the lowest-energy stationary solution, the relaxation of localized initial states, and the gravitational collapse of nonlocalized states. Our results show that contact interactions modify the density profile of the stationary solution and affect the onset of characteristic stages of gravitational collapse, particularly the shell-crossing event. In the (1+1) model, we confirm that relaxation does not converge to the lowest-energy stationary solution, even when local self-interactions are included. Taken together, local self-interactions play a relevant role in shaping the nonlinear dynamics of FDM and motivate further studies in higher-dimensional and cosmologically realistic settings.
}




\maketitle

\section{Introduction}\label{sec1}

Understanding the formation of structures remains one of the most intriguing puzzles in cosmology. Astronomical observations are currently consistent with the theory of a universe dominated by dark energy and cold dark matter, the so-called $\Lambda$CDM model. Its ability to reproduce large-scale observations makes it a highly reliable framework. However, this success is challenged on smaller scales, typically at distances below the separation between galaxies. In this regime, $\Lambda$CDM overpredicts the formation of structures, giving rise to inconsistencies such as the missing satellites and cusp-core problems, among others \cite{Bullock2025, deBlok2010}. Resolving these challenges necessitates improved knowledge of dark matter and dark energy.

Different strategies have been employed to deepen our understanding of dark matter. Among these, simulations across multiple length scales have played a central role in probing its nature \cite{Kuhlen2012}. N-body methods remain the standard tool for performing such simulations \cite{Springel2005}. Yet, the computational cost of increasing particle resolution to better capture the dynamics quickly becomes prohibitive. This limitation can be circumvented by adopting coarse-grained descriptions of the problem in terms of a continuous comoving phase-space density $f(\boldsymbol{x}, \boldsymbol{u})$, governed by the Vlasov-Poisson equation.

Cross-disciplinary approaches can provide fresh perspectives on such a challenging problem. In this context, fuzzy dark matter (FDM) constitutes an alternative model that may overcome some of the limitations of $\Lambda$CDM. Two complementary routes lead to this framework. The first postulates that dark matter consists of ultra-light particles ($m \sim 10^{-22}$ eV) that initially form a Bose-Einstein condensate \cite{Hu2000, Guth2015, Mocz2017, Elgamal2024}. The second exploits connections between the density field and smoothed phase-space representations, which allow the Vlasov-Poisson equations to be recast as Schr\"odinger-like equations \cite{Uhlemann2014, Schaller2014, Mocz2018}. Both approaches converge on the Schr\"odinger-Poisson equation \cite{Moroz:1998dh}, a model that also arises in diverse areas such as solid-state physics \cite{Su1991, Ram-Mohan2004}, nonlinear optics \cite{Bekenstein2015, Navarrete2017}, and ultracold atoms \cite{Sakaguchi2020, Ourabah2022}.

In the latter context, particle interactions play a crucial role in reproducing experimental results on Bose-Einstein condensation \cite{pethick2008bose, pitaevskii2016bose}. At very low energies the scattering process between particles is dominated by the s-wave contribution. Moreover, in this regime the scattering amplitude becomes independent of energy, approaching a constant value known as the scattering length. Consequently, the interaction between particles is fully characterized by this single parameter. Depending on the particle species, this interaction can be either attractive or repulsive, and in both cases  its principal effect is to modify the shape of the density profile \cite{castin2002bose}. This property is appealing in the dark matter setting, since it may provide a mechanism to alleviate the shortcomings of FDM.

The mass distribution of dark matter is expected to form a density profile characterized by a solitonic core. In three spatial dimensions, this soliton state acts as a dynamical attractor of gravitational collapse. This picture, however, changes significantly in reduced dimensionality. Although solitonic cores still form in one dimension \cite{Schulz2013}, the quasistationary state reached after gravitational collapse does not converge to the soliton \cite{Zimmermann2021}. As shown in that work, a closer correspondence with the (3+1)-dimensional case can be obtained by imposing strong harmonic confinement in the transverse directions.

One of the characteristic processes experienced by dark matter during collapse is known as shell-crossing. The fluid equations describing dark matter are commonly written in Lagrangian coordinates. In (1+1) dimensions, these equations become linear in such coordinates, allowing the fully nonlinear problem to be solved exactly. This linearization breaks down when the Lagrangian mapping becomes non-invertible, signaling the occurrence of shell-crossing. In (1+1) dimensions, this phenomenon is well understood, but its generalization to higher dimensions remains unclear. Perturbative treatments have been developed in quasi-one-dimensional systems \cite{Rampf2017}, while semiclassical approaches inspired by quantum mechanics offer complementary insights into the dynamics of dark matter \cite{Uhlemann2019}.

A related model appears in nonlinear optics, where the Schr\"odinger-Poisson equation with a contact interaction describes the nonlinear propagation of light beams, including Kerr effects \cite{Navarrete2017, Paredes2020}. The connection with cosmology has been highlighted in these studies, although contact interactions were not explicitly included. In addition, experimental setups have been proposed to investigate gravitational effects in laboratory optical systems \cite{Bekenstein2015}. Further work has shown that local self-interactions can give rise to soliton states with broader halos in the attractive case \cite{Painter2024}, while repulsive interactions lead to narrower halos \cite{Chen2021}. Moreover, in the analysis of collisions between gravitational structures \cite{Paredes2016}, repulsive interactions play a key role in stabilizing vortex states during binary soliton-vortex collisions \cite{Paredes2016, Nikolaieva2023}.

In this paper, we investigate the role of a contact interaction term in the dynamics of dark matter within the (1+1)-dimensional Schr\"odinger-Poisson framework. We explore both attractive and repulsive interactions through simulations in static and expanding universes. Our results show that contact interactions modify the shape of the lowest-energy stationary state and affect the onset of characteristic stages of gravitational collapse, with particular emphasis on the shell-crossing event. These findings may motivate further investigations into the role of contact interactions in more realistic cosmological settings, with potential implications for the formation of structure in the universe. 

\section{Theoretical background}\label{sec2}

The dynamics of a scalar field $\phi_\text{dm}$ of mass $m_\text{dm}$, describing dark matter in a gravitational field, can be derived from the Einstein-Hilbert action supplemented by the scalar-field contribution \cite{Suarez2014}. A matter contact interaction can be introduced by including a quartic local self-interaction term in the action \cite{Paredes2020},
\begin{equation}
S = \int d\hat{t} d^3 \hat{x} \sqrt{-g} \left[ \frac{c^3}{16 \pi G}(R - 2\Lambda) + \frac{\hbar^2}{2 c^2} (\partial \phi_\text{dm})^2 + \frac{m_\text{dm}^2}{2}\phi_\text{dm}^2  - \lambda_4 \frac{\hbar^3}{c^2} \phi_\text{dm}^4 \right],
\label{sec2ec1}
\end{equation} 
\review{where hatted symbols $\hat{t}$ and $\hat{x}$ denote dimensionful variables, which will later be related to dimensionless counterparts.  The parameter $\lambda_4$ is a dimensionless coupling that controls the quartic self-interaction term. In axionlike dark matter models, $\lambda_4$ is related to the mass $m_\text{dm}$ and the particle decay constant $f_a$ through the relation $\lambda_4 \sim m_\text{dm}^2 / f_a^2$ \cite{Guth2015, Levkov2017}.}

In an expanding, flat, and homogeneous universe slightly perturbed by the scalar field, the space-time element is given by the Friedmann-Lema\^itre-Robertson-Walker metric
\begin{equation}
ds^2 = -c^2 \left( 1 + 2 \frac{\hat{\Phi}}{c^2} \right) d\hat{t}^2 + \left( 1 - 2 \frac{\hat{\Phi}}{c^2} \right) a(\hat{t})^2 d\hat{x}_i d\hat{x}^i \,,
\label{sec2ec2} 
\end{equation} 
where $a(\hat{t})$ is the scale factor and $\hat{\Phi}$ is the Newtonian gravitational potential. Varying the action $\delta S$ with respect to $\phi_\text{dm}$ yields the scalar-field equation of motion, while variation with respect to the metric gives the Einstein field equations. After introducing a complex field $\hat{\psi}$, the scalar field can be rewritten as
\begin{equation}
\phi_\text{dm} = \sqrt{\frac{c}{2 m_\text{dm}}} \left( \hat{\psi} e^{-i m_\text{dm} c^2 \hat{t} / \hbar} + \hat{\psi}^\ast e^{i m_\text{dm} c^2 \hat{t} / \hbar} \right).
\label{sec2ec3}
\end{equation}
In the weak-gravity limit $\hat{\Phi} \ll c^2$ and the non-relativistic regime $\partial_{\hat{t}}^2 \hat{\psi} \ll (m_\text{dm} c^2 / \hbar) \partial_{\hat{t}} \hat{\psi}$, and after averaging out rapidly oscillating terms, the dynamics reduce to
\begin{equation}
i \hbar \partial_{\hat{t}} \hat{\psi} + \frac{3}{2} i \hbar \frac{\dot{a}}{a} \hat{\psi} = - \frac{\hbar^2}{2 m_\text{dm} a^2} \hat{\nabla}^2 \hat{\psi} + m_\text{dm} \hat{\Phi} \hat{\psi} - \frac{3 \lambda_4 \hbar^3}{m_\text{dm}^2 c}|\hat{\psi}|^2 \hat{\psi}.
\label{sec2ec4}
\end{equation}
Meanwhile, under the same approximations, the Einstein equations simplify to a Poisson equation sourced by the density contrast relative to the present-day total matter density $\rho_{m0}$,
\begin{equation}
\hat{\nabla}^2 \hat{\Phi} = 4 \pi G a^2 (|\hat{\psi}|^2 - \rho_{m0}).
\label{sec2ec5}
\end{equation}

Rescaling $\hat{\psi} \to a^{-3/2} \hat{\psi}$ then leads to a non-linear Schr\"odinger equation coupled to the Poisson equation for the gravitational potential,
\begin{subequations}
\begin{equation}
i \hbar \partial_{\hat{t}} \hat{\psi} = - \frac{\hbar^2}{2 m_\text{dm} a^2} \hat{\nabla}^2 \hat{\psi} + m_\text{dm} \hat{\Phi} \hat{\psi} - 3 \lambda_4 \frac{\hbar^3}{m_\text{dm}^2 c a^3} |\hat{\psi}|^2 \hat{\psi},
\label{sec2eq6a}
\end{equation}
\begin{equation}
\hat{\nabla}^2 \hat{\Phi} = \frac{4 \pi G}{a} \left(|\hat{\psi}|^2 - \rho_{m0} \right).
\label{sec2eq6b}
\end{equation}
\label{sec2eq6}
\end{subequations}

Equation (\ref{sec2eq6}) can be expressed in dimensionless form by introducing \cite{Zimmermann2019, Zimmermann2021, Schwersenz2024}
\[
\begin{array}{c c c}
\boldsymbol{x} = \left( \frac{m_\text{dm}}{\hbar} \right)^{1/2} \left[ \frac{3}{2} H_0^2 \Omega_{m0} \right]^{1/4} \hat{\boldsymbol{x}}, & &  dt = \frac{1}{a^2} \left[ \frac{3}{2} H_0^2 \Omega_{m0} \right]^{1/2} d\hat{t}, \\
\psi = \frac{\hat{\psi}}{\sqrt{\rho_{m0}}}, & & \Phi = a \frac{m_\text{dm}}{\hbar} \left[ \frac{3}{2} H_0^2 \Omega_{m0} \right]^{-1/2} \hat{\Phi},\\
\lambda = -3 \lambda_4 \frac{\hbar^2 \rho_m}{m_\text{dm}^2 c} \left[ \frac{3}{2} H_0^2 \Omega_{m0} \right]^{-1/2} & & ,
\end{array}
\]
where $H_0$ and $\Omega_{m0}$ denote the present-day Hubble constant and matter density parameter, respectively. This yields the Schr\"odinger-Poisson system,\begin{subequations}
\begin{equation}
i \partial_t \psi = \left( -\frac{1}{2} \nabla^2 + a \Phi + \frac{\lambda}{a} |\psi|^2 \right)\psi,
\label{sec2eq7a}
\end{equation}
\begin{equation}
\nabla^2 \Phi = |\psi|^2 - 1.
\label{sec2eq7b}
\end{equation}
\label{sec2eq7}
\end{subequations}

In the absence of a gravitational potential, Equation (\ref{sec2eq7}) reduces to the dimensionless Gross-Pitaevskii equation without an external potential. Here, two-body interactions are modeled as low-energy scattering processes characterized solely by the scattering length, whose sign determines whether interactions are attractive or repulsive. This correspondence motivates interpreting the quartic self-interaction term as a contact interaction.

Dimensional reduction from the (3+1)-dimensional system of Eq. (\ref{sec2eq7}) to a (1+1)-dimensional model can be implemented in different ways. A straightforward approach assumes a uniform matter distribution along the neglected dimensions, effectively replacing the Laplacian with a partial derivative in the remaining direction. Alternative reduction procedures include introducing an external confining potential in the transverse plane \cite{pethick2008bose, pitaevskii2016bose} or assuming angular symmetry \cite{baizakov2006matter}, the latter leading to a one-dimensional radial equation.

\section{Methods}\label{sec3}

To solve Equation (\ref{sec2eq7}) numerically in the domain $x \in \Omega = [-x_\text{max}, x_\text{max})$, for $t \in [0, t_\text{prop}]$, we discretize space and time as
\[
x_j = -x_\text{max} + j \Delta x, \quad j = 0, 1, \dots, N_x - 1, \quad \Delta x = 2 x_\text{max} / N_x,
\]
and 
\[
t_n = n \Delta t, \quad n = 0, 1, \dots, N_t, \quad \Delta t = t_\text{prop} / N_t,
\]
where $N_x$ and $N_t$ are even positive integers. The wavefunction evaluated at $(x_j, t_n)$ is denoted as $\psi(x_j, t_n) = \psi_j^n$. The equation is solved with the initial condition $\psi(x,0) = \psi_0(x)$ and periodic boundary conditions,
\begin{equation}
\begin{array}{c}
\psi(0, t) = \psi(L, t),\
\partial_x \psi(0, t) = \partial_x \psi(L, t), \quad L = 2x_\text{max}.
\end{array}
\label{sec3eq1}
\end{equation}

The Schr\"odinger-Poisson equation preserves the norm of the wavefunction $\psi$. In the context of dark matter dynamics, this property corresponds to conservation of mass during evolution:
\begin{equation}
M = \int_0^L dx |\psi(x,t)|^2 = L.
\label{sec3eq2}
\end{equation}
This election of the normalization condition ensures that the expectation value of the density contrast vanishes, $\langle \delta \rangle = \langle |\psi|^2 - 1 \rangle = 0$.

Equations (\ref{sec2eq7a}) and (\ref{sec2eq7b}) can be combined by writing the formal solution of the Poisson equation as a convolution,
\begin{equation}
\Phi(x, t) = \int_\Omega dx^\prime G_{1D}(x-x^\prime)|\psi(x^\prime)|^2,
\label{sec3eq3}
\end{equation}
where the kernel $G_{1D}(x, x^\prime)$ is given by \cite{Marshall2000, Zimmermann2020}
\begin{equation}
G_{1D}(x, x^\prime) = \frac{1}{L} \sum_{l>0} \left( -\frac{1}{k_l^2} \right) e^{-i k_l (x-x^\prime)}, \quad k_l = \frac{2\pi l}{L}, \ l = 0, \dots, N_x - 1.
\label{sec3eq4}
\end{equation}

Formally, the time-evolved wavefunction is obtained using the unitary time-evolution operator,
\[
\psi(x, t) = U(t)\psi(x, 0), \ U(t) = T \exp\left( -i \int_0^t dt^\prime H(t^\prime) \right),
\]
where $T$ is the time-ordering operator and $H(t) = H_\text{K}(t) + H_\text{R}(t)$ is the Hamiltonian, with $H_\text{K}$ and $H_\text{R}$ denoting the kinetic and potential parts. A single time step of size $\Delta t$ can thus be written as $\psi_j^{n+1} = U(\Delta t) \psi_j^{n}$.

The time-evolution operator can be approximated using a second-order Strang splitting \cite{glowinski2017splitting, May2021}:
\begin{equation}
U(\Delta t) = U_\text{K}\left( \tfrac{\Delta t}{2} \right) U_\text{R}\left( \Delta t \right) U_\text{K}\left( \tfrac{\Delta t}{2} \right) + \mathcal{O}(\Delta t^3),
\label{sec3eq5}
\end{equation}
where
\begin{subequations}
\begin{equation}
U_\text{K}(\Delta t) = \exp\left( - i H_\text{K} \Delta t \right), \quad H_\text{K} = -\tfrac{1}{2} \partial_x^2,
\label{sec3eq6a}
\end{equation}
\begin{equation}
U_\text{R}(\Delta t) = \exp\left( -i \int_t^{t+\Delta t} dt^\prime  H_\text{R}(t^\prime) \right), \quad H_\text{R}(t) = a(t)\Phi(x,t) + \frac{\lambda}{a(t)}|\psi(x,t)|^2.
\label{sec3eq6b}
\end{equation}
\label{sec3eq6}
\end{subequations}

The action of $U_\text{K}$ is most easily computed in momentum space, where it becomes diagonal:
\begin{equation}
U_\text{K}(\Delta t) \psi = \mathcal{F}^{-1} \left[ \exp\left(-i \tfrac{k^2}{2} \Delta t \right) \mathcal{F}\psi \right],
\label{sec3eq7}
\end{equation}
with $\mathcal{F}$ and $\mathcal{F}^{-1}$ denoting the Fourier transform and its inverse.

The operator $U_\text{R}$ requires a more careful treatment. The time dependence of $H_\text{R}$ arises from both the scale factor and the wavefunction. However, it can be shown that
\[
\partial_{t} |\psi_\text{R}|^2 = 0, \ \psi_\text{R} = U_\text{R} \psi,
\]
which implies that $|\psi(x, t)|^2 = |\psi(x, t_0)|^2$. Thus, the time dependence of $U_\text{R}$ reduces to the scale factor, and it suffices to compute its integral in Eq. (\ref{sec3eq6b}). Applying the midpoint method yields the following approximation for the time-evolution operator:
\begin{equation}
U_\text{R}(\Delta t) = \exp\left( -i a\left(t + \tfrac{\Delta t}{2}\right) \Phi(x, t_0) \Delta t + \frac{\lambda}{a\left(t + \tfrac{\Delta t}{2}\right)} |\psi(x, t_0)|^2 \Delta t \right).
\label{sec3eq8}
\end{equation}

The scale factor, which governs the expansion of the universe, obeys the Friedmann equation in a flat, radiation-free universe,
\begin{equation}
\left( \frac{\dot{a}}{a} \right)^2 = H_0^2 \left( \Omega_{m0} a^{-3} + \Omega_{\Lambda 0} \right).
\label{sec3eq9}
\end{equation}
In this work, we set the present-day density parameters to $\Omega_{m0} = 0.3$ and $\Omega_{\Lambda 0} = 1 - \Omega_{m0} = 0.7$. A closed-form expression for the scale factor is not available for arbitrary $\Omega_{m0}$, but it can be obtained numerically by following the procedure in \cite{Zimmermann2019}. This formulation enables the study of contact interaction effects in both a static universe ($a(t) = \text{const.}$) and an expanding universe.

\section{Results}\label{sec4}

\subsection{Ground states}\label{sec4subsec1}

We begin by examining the effect of the contact interaction on the ground state of the (1+1)-dimensional Schr\"odinger-Poisson equation, assuming a uniform matter distribution along the neglected dimensions. The ground state is defined as the stationary solution $\psi_\text{gs}(x)$ that minimizes the energy functional
\begin{equation}
E(\psi) = \int_\Omega dx \left[ \frac{1}{2} |\nabla \psi|^2 + \frac{1}{2} a \Phi |\psi|^2 + \frac{1}{2}\frac{\lambda}{a} |\psi|^4 \right].
\label{sec4eq1}
\end{equation}
It should be emphasized that $E(\psi)$ is conserved only in a static universe; therefore, in this section we consider a constant scale factor. Several approaches exist for solving this minimization problem \cite{Bao2013Review}. Here, we adopt the well-known imaginary time propagation method.

\review{QCD axions models constrain the decay constant to lie in the range $10^9 - 10^{12}$ GeV. For ultralight bosons with masses $m \sim 10^{-19}$ eV, this constraint translates into values of the self-interaction parameter $\lambda$ that fall within the range explored in our simulations.}

\begin{figure}
\begin{center}
\includegraphics[scale=0.55]{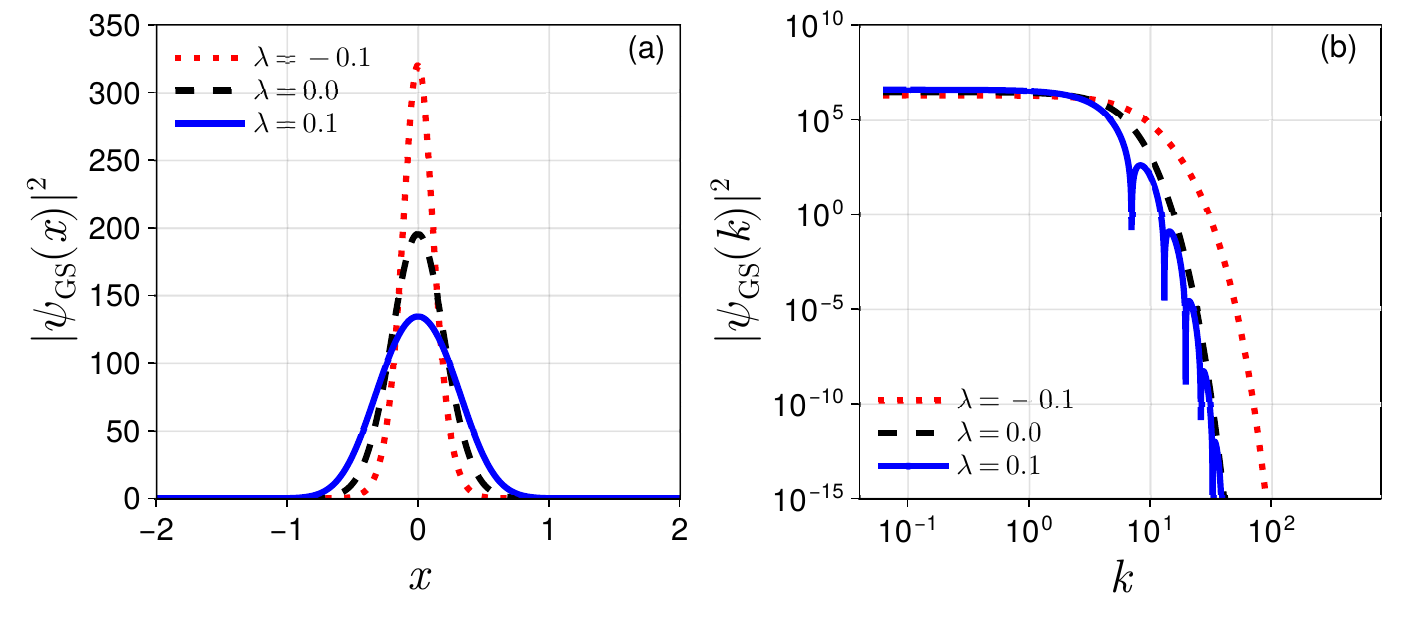}
\end{center}
\caption{Ground states obtained with the imaginary-time propagation method for attractive and repulsive contact interactions at fixed mass $M = 100$ in a static universe ($a = 1$). (a) Density profiles for different values of $\lambda$, including the purely gravitational case ($\lambda = 0$) for reference. (b) Ground-state spectra for varying contact interaction strengths.}
\label{fig1}
\end{figure}

Figure \ref{fig1} shows the ground state obtained for different values of the contact interaction strength. As illustrated in panel (a), the contact interaction modifies the shape of the ground state, making it broader or narrower depending on whether it is repulsive or attractive. The stronger the interaction, the more pronounced the change in the ground-state profile. Our codes and data used to create Figure \ref{fig1} are publicly available, see \cite{RodriguezVillalbaDS}.

This trend is consistent with the behavior observed in higher dimensions. The influence of local interactions on density profiles has been examined in the (3+1)-dimensional system \cite{Chavanis2011_p1, Chavanis2011_p2, Chen2021, Painter2024}. Attractive contact interactions lead to narrower profiles with higher central densities, whereas repulsive interactions produce broader profiles with lower central densities.

The broadening of the density profile corresponds to a narrowing of the spectrum, as shown in panel (b) of Figure \ref{fig1}. In the attractive and non-interacting cases the spectra remain smooth, whereas in the repulsive case additional peaks manifest.  These structures encode information about the shape and smoothness of the density profile. A ground-state density profile with sharper edges leads to more pronounced peaks in the spectrum.

\begin{figure}
\begin{center}
\includegraphics[scale=0.5]{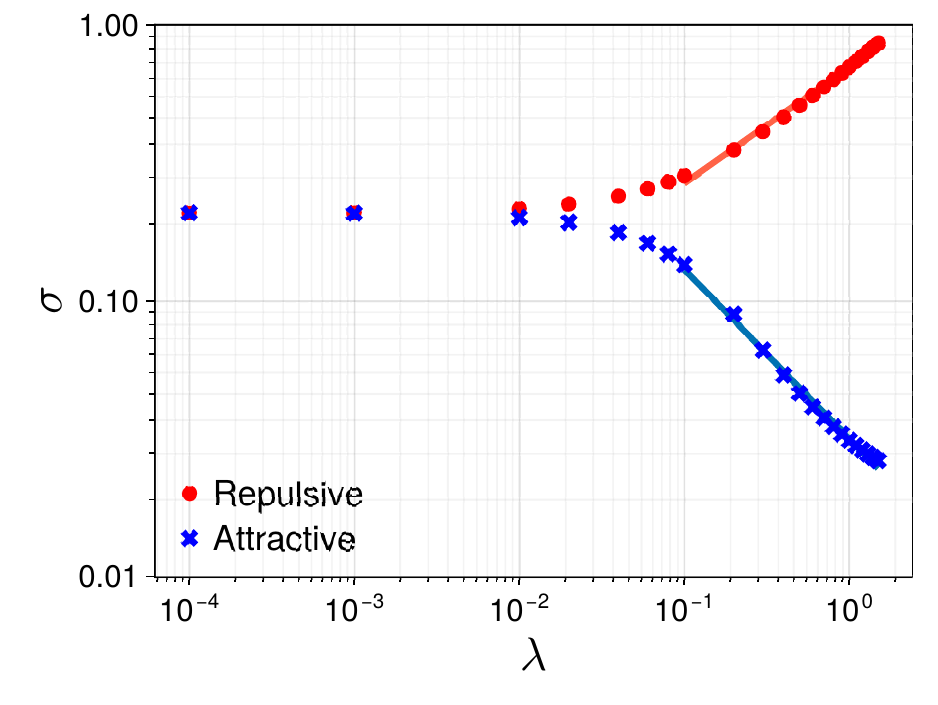}
\end{center}
\caption{Ground-state width as a function of the contact interaction strength $\lambda$ for $a = 1$ and mass $M = 100$. Solid lines indicate the points that fit well to a power-law function.}
\label{fig2}
\end{figure}

To further quantify the impact of the contact interaction on the ground-state shape, we plot the standard deviation as a function of the interaction strength in Figure \ref{fig2}. To compute this quantity we interpret $|\psi|^2$ as a probability density and evaluate it on our grid using the expression
\[
\sigma = \sqrt{\frac{\sum_j (x_j - \mu)^2 |\psi_j|^2 \Delta x}{\sum_j |\psi_j|^2 \Delta x}},
\]
where $\mu = 0$ is the mean value of $|\psi|^2$ for all the simulations considered. For small $\lambda$, the ground-state width is essentially the same for both attractive and repulsive interactions. As the interaction strength increases, however, their behaviors diverge: the ground-state width grows in the repulsive case and shrinks in the attractive case. For sufficiently large interaction strengths, the standard deviation follows a power law, with $\sigma \propto \lambda^{0.42}$ for repulsive interactions and $\sigma \propto |\lambda|^{-0.15}$ for attractive interactions.

\review{These ground states are characterized by a balance among kinetic, gravitational, and contact interaction terms. For sufficiently large contact interaction strengths, one of these contributions becomes negligible with respect to the other two, which then mainly determine the balance.  
}

\review{
For repulsive contact interactions, the kinetic contribution becomes negligible. In the context of cold-atom physics, this regime is known as the Thomas-Fermi approximation. In this limit, the density profiles are characterized by a length scale $R \sim \lambda^{0.5}$ \cite{Chavanis2011_p1}, in close agreement with our power-law fit. In contrast, for attractive contact interactions, the numerical verification of the balance shows that the kinetic and contact terms dominate. As the interaction strength increases, the density profiles become progressively narrower. These features resemble Bose-Einstein condensates with attractive contact interactions, for which stationary solutions--known as bright solitons--are characterized by a length scale $R \sim |\lambda|^{-0.5}$ \cite{pethick2008bose, pitaevskii2016bose}. Quantitative discrepancies are expected 
due to the additional presence of the gravitational potential and because sharply peaked density profiles produce steep gradients, which hinder the numerical verification of the balance condition.
}

In summary, an attractive contact interaction amplifies the nonlinear focusing effect of gravity, while a repulsive interaction enhances the dispersive effect of quantum pressure. The asymmetry observed in the scaling of the ground-state width for attractive versus repulsive interactions reflects the inherently asymmetric contribution of the contact term.

\subsection{Relaxation}\label{sec4subsec2}

Following the discussion in \cite{Zimmermann2021}, we now consider the relaxation process of an initial Gaussian wavepacket,
\[
\psi_0(x) = N e^{-\frac{x^2}{2 \sigma^2}}
\]
where $N$ is a normalization constant. A detailed analysis of the relaxation mechanisms and the different evolution phases experienced by the wavepacket can be found in \cite{Zimmermann2021}. Here, we focus on assessing whether the ground state--obtained following the procedure outlined in Section \ref{sec4subsec1}--serves as a dynamical attractor of the relaxation process when the contact interaction is included.

To this end, we simulate the time evolution of this initial localized state up to $t_\text{prop} = 100$, a period extending beyond the time at which the Gaussian wavepacket relaxes in the pure gravitational scenario. The standard deviation is given by $\sigma = b k_\text{J}^{-1}$ where $b$ is a positive constant and $k_\text{J}^{-1}$ defines the Jeans length $\lambda_\text{J} = 2 \pi / k_\text{J}$. This length scale marks the point where quantum pressure compensates gravitational attraction, thereby preventing the collapse of density fluctuations. Local self-interactions modify the Jeans length: it becomes larger in the repulsive case and shorter in the attractive one, relative to the pure gravitational scenario \cite{Chavanis2011_p1}. For consistency, we select the standard deviation based on the pure gravitational Jeans length, which provides an intermediate scale between those corresponding to attractive and repulsive contact interactions.

By adjusting the value of the constant $b$, we modify the spatial width of the initial state relative to the Jeans length, thereby controlling the instabilities present at the start of the evolution. The length of the simulation domain is chosen as 
$L = (180/b) \sigma \approx 127$. To minimize boundary effects in the simulations, small values of $b$ are therefore preferred.

\begin{figure}
\begin{center}
\includegraphics[scale=0.55]{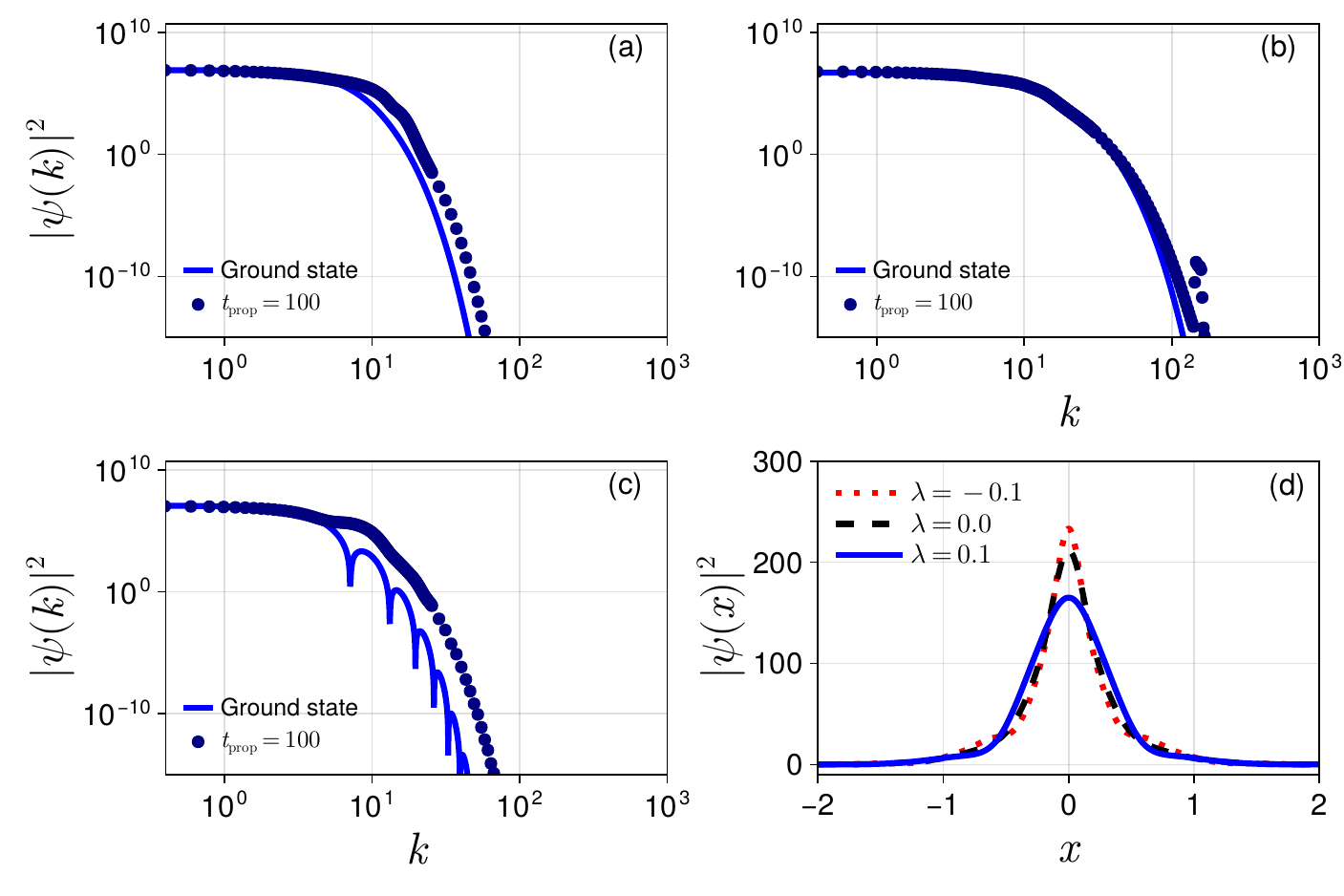}
\end{center}
\caption{Comparison between the mean quasistationary states obtained after the relaxation process and the corresponding ground states for (a) no contact interaction, (b) attractive contact interaction, and (c) repulsive contact interaction. Panel (d) shows the mean densities: the peak is more pronounced in the attractive case compared with the pure gravitational and repulsive cases, while the spatial width is larger for the repulsive contact interaction. The mean spectra and densities were obtained by first propagating the initial state up to $t = 90$; during the remaining time interval, 500 wavefunctions were saved. Finally, the densities and spectra were computed and averaged.}
\label{fig3}
\end{figure}

Figure \ref{fig3} compares the relaxed state obtained after time evolution with the corresponding ground state for three cases: no contact interaction, attractive contact interaction, and repulsive contact interaction. Further details of the simulations and source codes to produce this figure can be found in \cite{RodriguezVillalbaDS}. The standard deviation of the initial Gaussian state in these simulations was $\sigma = k_\text{J}^{-1}$. Although the relaxed state exhibits the same qualitative influence of the contact interaction on its shape as observed for the ground state, there is no quantitative agreement between the two for any value of the interaction strength. The relaxed state consistently shows a broader spectrum than the ground state, with smaller discrepancies in the attractive case and larger ones in the repulsive case. The discrepancy increases for larger values of the standard deviation of the initial state and is more pronounced for the repulsive and pure gravitational cases compared to the attractive case. 

The proximity of the relaxed state to the ground state can be quantified by computing the fidelity,
\begin{equation}
F(t) = \frac{\abs{\ip{\psi_\text{gs}(x)}{\psi(x,t)}}^2}{L^2},
\label{sec4eq2}
\end{equation}
throughout the evolution. This quantity is defined such that $F=1$ corresponds to complete agreement between the states. The fidelity exhibits an oscillatory pattern around a central value that depends on the instabilities present in the initial state. For larger values of the standard deviation, the fidelity displays more pronounced oscillations around smaller mean values. Conversely, smaller standard deviations yield smaller oscillations around higher fidelities, indicating a closer resemblance between the relaxed state and the ground state.

When comparing the three scenarios, repulsive contact interactions yield more stable states, characterized by smaller deviations of the fidelity around larger mean values, whereas attractive interactions lead to more unstable states, with larger deviations around smaller mean values. This behavior can be interpreted in terms of the Jeans length. In the linear regime, perturbations remain stable for $k > k_\text{J}$ and grow for $k < k_\text{J}$ \cite{Chavanis2011_p1}. Consequently, a smaller Jeans length--as  occurs for attractive interactions--is associated with a large number of unstable modes. Finally, it is worth noting that the fidelity primarily quantifies similarity in the core region and does not capture differences in the halos.

In summary, although the relaxation process produces a localized structure that displays the same qualitative dependence on the contact interaction as the ground state, our results indicate that the system does not converge to the ground state in the (1+1)-dimensional model. This extends the findings of \cite{Zimmermann2021} to include the presence of contact interactions.

\subsection{Gravitational collapse in an expanding universe}\label{sec4subsec3}

In this section, we study the dynamics of wavepackets in an expanding universe. To this end, we performed simulations for both attractive and repulsive contact interactions while accounting for the time dependence of the scale factor.

A common test for CDM simulations is to investigate the gravitational collapse of a sine wave in an expanding universe \cite{Zimmermann2019, Kopp2017}:
\[
\psi_0(x) = \sqrt{A \cos(\frac{2 \pi}{L} x) + 1}.
\]
This state provides a good model of density fluctuations on small spatial domains \cite{Schwersenz2022}. In our simulations, we set the amplitude to $A = 0.1$.

We study the time evolution from a past epoch to the present day, where the scale factor is normalized to $a = 1$. The initial time is most conveniently specified in terms of the redshift $z$, with $z = 0$ corresponding to the present. The scale factor is related to the redshift by $z + 1 = 1/a$. Our simulations span from $z = 500$ to $z = 0$, corresponding to $t = 0$ and $t = 52$ in dimensionless time, respectively.

CDM models assume extremely weak interactions between particles, making them effectively collisionless on cosmic scales. Within this framework, introducing a contact interaction may appear inconsistent. However, the last term in Eq. (\ref{sec2eq7a}) represents a contact interaction whose strength decreases over time due to cosmic expansion, becoming negligible at present.

Contact interactions are expected to modify the dynamics of matter during gravitational collapse, potentially addressing inconsistencies in the predicted structure formation of FDM models \cite{Painter2024}. In CDM, the collisionless nature of the particles allows their trajectories to overlap, meaning different initial positions can lead to the same final position. The region before trajectory crossing is called the single-stream region, while the region afterward is the multi-stream region. The first occurrence of multistreaming is known as the first shell-crossing event \cite{Gough2022}, a key phase of gravitational collapse. For this reason, we use it to evaluate the impact of contact interactions.

Single- and multi-stream regions are separated by caustics \cite{berry1980iv}, where the density formally diverges. In one dimension, this density can be expressed exactly in terms of the Jacobian of the Lagrangian map, with the divergence corresponding to the vanishing of the Jacobian, thereby marking the onset of shell-crossing. In higher dimensions, the behavior becomes more complex \cite{Rampf2017}, and the influence of additional interactions on shell crossing is not yet fully understood.

In our framework, dark matter is modeled by a complex wavefunction. Thus, trajectory overlap in CDM corresponds to interference of the wavefunction with itself. Regions of maximum constructive interference can therefore be associated with caustics. This correspondence has been highlighted in semianalytical studies of shell crossing based on semiclassical approximations \cite{Gough2022}.

\begin{figure}
\begin{center}
\includegraphics[width=\textwidth, height=10cm]{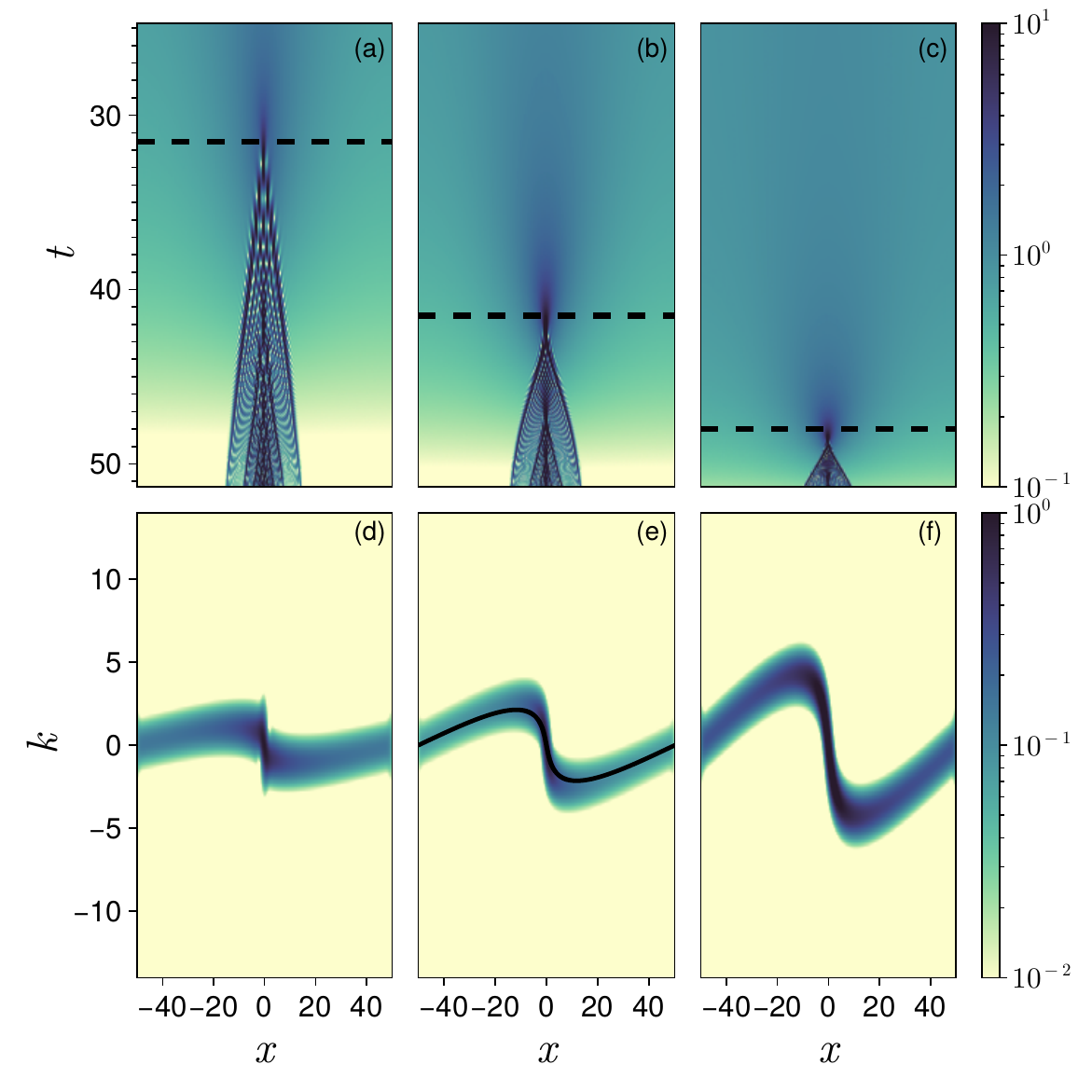}
\end{center}
\caption{Spatiotemporal representation of the wavepacket for (a) attractive, (b) no, and (c) repulsive contact interaction. The initial interaction strength is $\abs{\lambda/a_\text{ini}} \approx 1.25$. The corresponding shell-crossing times are $t_\text{sc} = 31.5$ ($z_\text{sc} = 89.6$), $t_\text{sc} = 41.5$ ($z_\text{sc} = 28.6$), and $t_\text{sc} = 48.0$ ($z_\text{sc} = 6.7$), respectively. Panels (d)-(f) display the associated Husimi functions at shell crossing. For equal resolution in both directions, the parameter was set to $\sigma_x = 1/\sqrt{2}$. In panel (e), the classical solution is shown as a black solid line.}
\label{fig4}
\end{figure}

A spatiotemporal representation of the wavefunction provides a practical way to identify the first shell-crossing event. Figure \ref{fig4} (panels (a) and (c)) shows the time evolution for both attractive and repulsive contact interactions. The horizontal dashed line marks the onset of shell-crossing, identified as the moment when self-interference first appears. The onset occurs earlier in the attractive case and later in the repulsive case, relative to the pure gravitational scenario.

To further confirm the occurrence of shell-crossing, we compute the Husimi representation of the wavepacket, defined as $f_\text{H}(t, x, k) = \abs{Q_\text{H}(t, x, k)}^2$ \cite{wimberger2014nonlinear, Kopp2017}, with
\[
\begin{split}
& Q_\text{H}(t, x, k) = \int dx^\prime K_\text{H}(x, x^\prime, k) \psi(t, x^\prime),\\
& K_\text{H}(x, x^\prime, k) = \frac{\exp[-\frac{(x-x^\prime)^2}{4\sigma_x^2}-\frac{i}{\hbar} k x^\prime]}{(2 \pi \hbar)^{1/2} (2 \pi \sigma_x^2)^{1/4}}. 
\end{split}
\]
The parameter $\sigma_x$ controls the resolution of the phase-space density $f_\text{H}(t, x, k)$ in the $x$ and $k$ directions, with $\sigma_k = \hbar / (2\sigma_x)$. At the onset of shell crossing, regions of the Husimi function become perpendicular to the spatial axis, signaling the transition from a single-stream (single-valued) to a multi-stream (multi-valued) regime. In the repulsive case, the Husimi function exhibits higher $k$-mode components, indicating faster motion of matter streams during shell crossing. For the pure gravitational case, we have included the classical solution obtained by adapting the approach of \cite{pietroni2018structure} to the Friedmann equation adopted in this work. The close agreement underscores the role of the classical solution as a backbone of the quantum description.

\begin{figure}
\begin{center}
\includegraphics[scale=0.5]{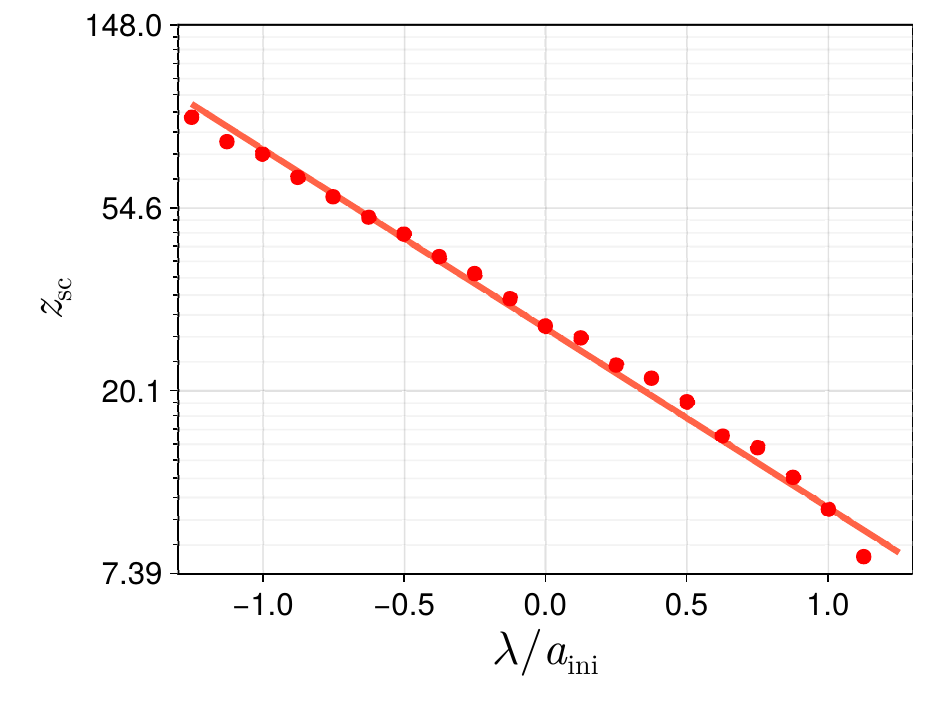}
\end{center}
\caption{Onset of the first shell crossing as a function of the contact interaction strength. Negative values on the horizontal axis correspond to attractive interactions, while positive values correspond to repulsive interactions. The solid line shows the exponential fit.}
\label{fig5}
\end{figure}

The onset of the first shell crossing depends on both the strength and the type of contact interaction. Figure \ref{fig5} shows the redshift of shell crossing as a function of interaction strength. The data can be fitted by the exponential law
\[
z_\text{sc} = z_0 e^{-0.97 \frac{\lambda}{a_\text{ini}}},
\]
where $z_0 = 28.6$ is the shell-crossing redshift in the pure gravitational case and $a_\text{ini}$ is the initial scale factor.

According to this relation, attractive contact interactions advance the occurrence of shell crossing by enhancing gravitational attraction, while repulsive contact interactions delay it by reinforcing quantum pressure against gravity.

\section{Conclusion}\label{sec5}

In this work, we studied the impact of a contact interaction term on the dynamics of dark matter modeled by the Schr\"odinger-Poisson equation in (1+1) dimensions. We performed simulations with both attractive and repulsive contact interactions, considering static as well as expanding universes. Three scenarios were analyzed: (i) the properties of the ground state (the lowest-energy stationary state), (ii) the quasistationary state emerging from a localized initial wavepacket in a static universe, and (iii) the gravitational collapse of a nonlocalized initial state in an expanding universe. The wave-based nature of our framework allowed us to use spatiotemporal and phase-space representations, together with quantum measures such as fidelity, to characterize the key dynamical processes.

We found that the inclusion of contact interactions modifies the ground-state shape in agreement with previous results for the (3+1) Schr\"odinger-Poisson system. Our (1+1) analysis extends these results by systematically exploring both attractive and repulsive cases. We also observed a discrepancy between the final state after relaxation and the ground state, confirming that in contrast to the (3+1) model, the ground state does not act as a dynamical attractor in the (1+1) setting. Furthermore, we showed that in an expanding universe the contact interaction term can advance or delay the stages of gravitational collapse, with attractive interactions enhancing gravitational focusing and repulsive ones strengthening quantum pressure. Although shown within a simplified cosmological framework, this effect points to a potential mechanism for addressing discrepancies between fuzzy dark matter predictions and small-scale observations of structure formation.

\review{We have studied shell crossing in one dimension (1D), which is considerably simpler than its three-dimensional (3D) counterpart. In the case $\lambda = 0$, it is well established that the linear (Zel'dovich) approximation is exact in 1D up to the onset of shell crossing, and that (semi-)analytical treatments exist to describe both the evolution beyond the first shell crossing \cite{Colombi2015, Taruya2017, Rampf2021} and the asymptotic late-time \cite{pietroni2018structure, Chen2020} behavior. Extending these 1D results to fully three-dimensional dynamics, however, involves substantial technical and conceptual challenges. In 3D, the geometry of multistream regions is far more complex, and shell crossing typically occurs along different spatial directions at different times. Nonetheless, in the pre-collapse regime, the possibility of at least qualitatively extending the 1D insights to 3D rests on a well-established framework, namely Zel'dovich's pancake model of structure formation. Within Lagrangian perturbation theory, this model predicts anisotropic collapse: matter first contracts along a single principal axis, forming effectively one-dimensional, sheet-like ``pancakes'', which subsequently collapse along additional axes to generate filaments and halos. Consequently, prior to shell crossing, the 3D dynamics is, to a good approximation, effectively one-dimensional and thus amenable to perturbative analysis. We therefore expect that our findings should qualitatively carry over to the 3D case, including the $\lambda$-dependence of the shell-crossing redshift. For a recent treatment of the 3D problem in terms of an effective 1D system, see Ref. \cite{Saga2025}}.

Our results highlight both the potential and the limitations of lower-dimensional approaches. A possible extension to our (1+1)-dimensional analysis could be to incorporate local interactions into the framework of \cite{pietroni2018structure}, either through a classical treatment or within a semiclassical description. While such (1+1)-dimensional simulations provide valuable insight into the qualitative role of contact interactions, extensions to higher dimensions and more realistic cosmological settings are necessary to fully assess their impact. Future work could explore such higher-dimensional generalizations, investigate the interplay with additional self-interaction terms, and test whether the delay or advancement of collapse stages persists in more complex scenarios.

\section*{Declarations}

\subsection*{Acknowledgements}

We are very grateful to Tim Zimmermann for his valuable contributions to the early stage of our work on the Schr\"odinger-Poisson model and to Luca Amendola for critical discussions.

\subsection*{Funding}

O. Rodr\'iguez-Villalba, I. Saychenko and S. Wimberger acknowledge funding by Q-DYNAMO (EU HORIZON-MSCA-2022-SE-01) with project No. 101131418, and by the National Recovery and Resilience Plan, through PRIN 2022 project "Quantum Atomic Mixtures: Droplets, Topological Structures, and Vortices", project No. 20227JNCWW, CUP D53D23002700006, and through Mission 4 Component 2 Investment 1.3, Call for tender No. 341 of 15/3/2022 of Italian MUR funded by NextGenerationEU, with project No. PE0000023, Concession Decree No. 1564 of 11/10/2022 adopted by MUR, CUP D93C22000940001, Project title "National Quantum Science and Technology Institute" (NQSTI). M. Pietroni acknowledges support by the MIUR Progetti di Ricerca di Rilevante Interesse Nazionale (PRIN) Bando 2022 - grant 20228RMX4A, funded by the European Union - Next generation EU, Mission 4, Component 1, CUP C53D23000940006. 

\subsection*{Competing interests}

The authors declare no competing financial interests.

\subsection*{Author contributions}

All authors contributed to the design of the research and the writing of the manuscript. ORV produced the data. SW supervised the work.

\subsection*{Data availability}

Data and source codes to produce the plots included in this paper can be found in \cite{RodriguezVillalbaDS}.


\end{document}